\documentclass[12pt]{iopart}

\expandafter\let\csname equation*\endcsname\relax
\expandafter\let\csname endequation*\endcsname\relax

\usepackage[dvips]{graphics}
\usepackage{amsmath,color}
\usepackage{amsfonts}
\usepackage{soul}

\usepackage{amssymb}
\usepackage{amstext}

\usepackage{braket}
\usepackage{bbold}
\usepackage{mathtools}
\usepackage{gensymb}

\newcommand{\ignore}[1]{}

\makeatletter
\newcommand*{\defeq}{\mathrel{\rlap{%
                     \raisebox{0.3ex}{$\m@th\cdot$}}%
                     \raisebox{-0.3ex}{$\m@th\cdot$}}%
                     =}
\makeatother

\begin{document}

\title{\textbf{Smallest state spaces for which bipartite entangled quantum states are separable}}
\author{Hussain Anwar$^{1}$, Sania Jevtic$^1$, Oliver Rudolph$^2$ and Shashank Virmani$^1$}

\address{$^1$ Department of Mathematical Sciences, Brunel University, Uxbridge, Middlesex UB8~3PH, United Kingdom}

\address{$^2$ Leonardo da Vinci Gymnasium, Im Spitzerfeld 25, 69151 Neckargem\"und, Germany
\& Hector-Seminar, Waldhoferstr.~100, 69123 Heidelberg, Germany.}

\ead{hussain.anwar@brunel.ac.uk}


\begin{abstract}
According to usual definitions, entangled states cannot be given a separable decomposition in terms of products of local density operators. If we relax the requirement that the local operators be positive, then an entangled quantum state may admit a separable decomposition in terms of more general sets of single-system operators. This form of separability can be used to construct classical models and simulation methods when only restricted set of measurements are available. With these motivations in mind, we ask what are the smallest sets of local operators such that a pure bipartite entangled quantum state becomes separable? We find that in the case of maximally entangled states there are many inequivalent solutions, including for example the sets of phase point operators that arise in the study of discrete Wigner functions. We therefore provide a new way of interpreting these operators, and more generally, provide an alternative method for constructing local hidden variable models for entangled quantum states under subsets of quantum measurements.
\end{abstract}

\submitto{\NJP}

\maketitle

\section{Overview}

Entangled quantum systems are a powerful resource for many quantum information processes because of their non-classical correlations. The joint statistics of separated measurements cannot be described by probability distributions over local hidden variables \cite{B64}, and in some quantum computational models it is known that entanglement is a prerequisite for non-classical computation \cite{nonclassical}.

However, if the local measurements made on a quantum system are restricted, as is the case in many settings, then the measurement statistics can admit a local description for the system, even when it requires an entangled-state description for unrestricted measurements. For instance, it has been known since the time of Bell that a two-qubit EPR pair has a local hidden variable (LHV) model for the Pauli measurements \cite{Bell}. This fact can be reinterpreted as a statement that the EPR pair is separable (i.e. non-entangled) with respect to a more general set of single-system operators, consisting of cubes of Bloch vectors that enclose the usual Bloch sphere \cite{Shash1}. While the operators that correspond to these `cube' Bloch vectors are not always physical, they can be considered as valid state descriptions if measurements are restricted to the Pauli operators.

The investigation of such non-quantum spaces resides in the study of generalized probabilistic theories \cite{Boxworld,Barnum}. This field of research considers theories more general than quantum theory by describing single- and multi-particle systems in terms of tables of probabilities for measurement outcomes, under various natural constraints such as not allowing instant signaling. In principle such theories do not necessarily have an underlying structure in terms of Hilbert spaces and operators, and can exhibit correlations that are more powerful than quantum theory.

In this paper we will build upon these ideas to construct separable descriptions for entangled quantum states. In particular, we will set out to find the {\it smallest} local state spaces such that a given quantum entangled state can be considered separable. The reason for doing this is that, as we discuss later, such state spaces will typically be the most useful for constructing local descriptions of entangled states in various settings. Unlike the more general formalism of generalized probabilistic theories, however, the local state spaces that we consider still have some quantum structure in that they are sets of operators with the same dimensions as the density operators \cite{OperatorGPT}. Indeed, the only correlations that we consider arise from quantum systems.

The structure of this paper is as follows. In the next two sections we set up notation and define more precisely the various problems that we consider. In section \ref{summ} we summarise the methods that we use. In section \ref{connections} we discuss connections between our problems and the study of cross norm measure of entanglement \cite{OR1,OR2}. In sections \ref{smallest}--\ref{prob3} we construct solutions to our problems for the case of maximally entangled states, and in section \ref{conclusions} we conclude. 

\section{Generalized separability}

In the conventional quantum description of entanglement, a quantum state of two or more particles is said to be entangled if it cannot be written as a probabilistic mixture of products of single particle quantum states \cite{W89}. The textbook example of an entangled quantum state for two $d$-level quantum particles is the maximally entangled state, denoted here by $ \ket{\phi_d} $, which by a suitable local basis choice can be written in the form:
\begin{eqnarray}
\ket{\phi_d} = {1 \over \sqrt{d}} \sum_{j=0}^{d-1} \ket{jj}.
\end{eqnarray}
It is well known that this state cannot be written in the form of a separable state,
\begin{eqnarray}
\ket{\phi_d}\bra{\phi_d} \neq \sum_j p_j \rho^A_j \otimes \rho^B_j,
\end{eqnarray}
where $\rho^A_j$ and $\rho^B_j$ are drawn from the sets $\mathcal{Q}_A$ and $\mathcal Q_B$ of quantum states on each subsystem, and $p_j$ forms a probability distribution. However, if we relax the restriction that the local operators $\rho^A_j$ and $\rho^B_j$ be drawn from $\mathcal Q_A$ and $\mathcal Q_B$, then we can indeed find separable decompositions for $\ket{\phi_d}$. If two convex sets of operators $\mathcal V_A$ and $\mathcal V_B$ are such that a given quantum state $\Psi$ can be written as:
\begin{eqnarray}
\Psi = \sum_j p_j \rho^A_j \otimes \rho^B_j,   \label{gensep}
\end{eqnarray}
where $\rho^A_j \in \mathcal V_A$ and $\rho^B_j \in \mathcal V_B$, and $p_j$ forms a probability distribution, then we will say that $\Psi$ is $\mathcal V$-separable, where $\mathcal V$ denotes the pair of sets $(\mathcal V_A,\mathcal V_B)$.

This generalized notion of separability \cite{Barnum} can adopt practical significance if the local operators in the separable decomposition exhibit some form of what can be referred to as {\it generalized positivity}. In the usual study of quantum separability, the notion of positivity amounts to matrix-positivity of the local operators so that they can correspond to density matrices. However, in other contexts, alternative notions of positivity may be useful. One example of such an alternative notion is positivity defined with respect to a subset of quantum measurements. This can be defined using the notion of a {\it dual}. Give a set $\mathcal{X}$ of operators, the dual of the set is defined as the set of operators satisfying $\mathcal{X}^*:=\left\{H\mid 0\leq\tr\left(H G^{\dag}\right) \forall G \in \mathcal{X}\right\}$. Due to its connection with the Born rule, this definition can be applied to define sets of operators that are `positive' for subsets of of quantum measurements. 

Consider a positive-operator valued measurement (POVM) performed on a single quantum particle:
\begin{eqnarray}
\mathcal{M}=\left\{M_i\mid \sum\nolimits_i M_i = \mathbb{1},~M_i \geq 0 \right\}. \nonumber
\end{eqnarray}
We define the {\it dual} set of this measurement, denoted by $\mathcal{M}^*$,  as the set of operators that give positive values under the Born rule:
\begin{eqnarray}
\mathcal{M}^*\defeq\left\{X\mid ~ 0\leq\tr\left(X M_i\right),~\forall~M_i\in \mathcal{M}\right\}. \nonumber
\end{eqnarray}
(Notice that a POVM consists of Hermitian operators, so it is irrelevant whether we place the dagger ``$^{\dagger}$" on the $M_i$). In this context we say that the elements of $\mathcal{M}^*$ are {\it positive} with respect to $\mathcal{M}$, or more concisely $\mathcal{M}$--positive. Related definitions appear in the context of generalized probabilistic theories \cite{Boxworld}. Some authors define the dual as consisting only of operators with restrictions on normalisation. However, we do not do that here as in one application we will consider the construction of local hidden variable models, and in those contexts it is more natural to not impose that restriction.

This definition can be extended to collections of quantum measurements. If instead of a single POVM we consider a collection $\mathcal F$ of POVMs on the same particle, then we define the dual $\mathcal F^*$ as the intersection of sets of operators that give valid probability distributions for all of the measurements in the collection:
\begin{eqnarray}
\mathcal{F}^*\defeq\bigcap_{\mathcal{M}\in\mathcal{F}} \mathcal{M}^*, \nonumber
\end{eqnarray}
and we describe the elements of $\mathcal{F}^*$ as being $\mathcal{F}$-positive. Note that to be positive with respect to any set of measurements, an operator must have a non-negative trace. The full dual set $\mathcal{F}^*$ forms a {\it convex cone}, i.e. if $X_1, X_2 \in \mathcal{F}^*$, then so is $x_1 X_1 + x_2 X_2$ for any coefficients $x_1,x_2\geq 0$. 

If the local operators appearing in a generalized separable decomposition are positive with respect to the measurements on each particle, then this can help to provide classical descriptions for the state of a bipartite (or for that matter multipartite) system. If, for example, the sets $\mathcal V_A$ and $\mathcal V_B$ are subsets of the cones of $\mathcal F_A$ and $\mathcal F_B$ positive operators on particles $A$ and $B$, respectively, and if a quantum state $\Psi$ is $\mathcal V$-separable, then the separable decomposition supplies a local hidden variable model for measurements from $\mathcal F_A$ and $\mathcal F_B$ made on $\Psi$ \cite{foot98}. Moreover, in some cases such separable decompositions can help to efficiently classically simulate quantum entangled systems under $\mathcal F$, see for instance \cite{Shash1,Shash2,AJRV2}.

With such applications in mind, our goal in this work will be to try to identify the {\it smallest} choices for $\mathcal V_A$ and $\mathcal V_B$ such that the maximally entangled state $\ket{\phi_d}$ is $\mathcal V$-separable. While we make our definitions of `small' more precise in the rest of the paper, the motivation for this problem is that for most reasonable forms of generalised positivity, the positivity of a set of operators guarantees the positivity of any of its subsets, and this means that identifying smallest state spaces under which a given state is separable will lead us to `more positive' separable descriptions of the state. Consider, for instance, two sets $\mathcal V_A \subset \mathcal W_A$ for which we know that a quantum state $\Psi$ is $\mathcal (\mathcal{V}_A,\mathcal{V}_B)$-separable (and hence also $\mathcal (\mathcal{W}_A,\mathcal{V}_B)$-separable). Then, the set of measurements for which $ \mathcal V_A $ is positive cannot be smaller than the set of measurements for which $ \mathcal W_A $ is positive. This in turn implies that $(\mathcal{V}_A,\mathcal{V}_B)$-separability supplies a LHV model for a no-smaller class of measurements than $\mathcal (\mathcal{W}_A,\mathcal{V}_B)$-separability.

While there can be many definitions of `size' for the sets $\mathcal V_A$ and $\mathcal V_B$, we will choose definitions that enable us to make analytical progress, and in the process identify choices of $\mathcal V_A$ and $\mathcal V_B$ that are the `smallest' possible---in that no strict subsets of them can be chosen while keeping $\ket{\phi_d}$ separable. However, we will see that there are many such inequivalent solutions. Amongst them is the set of phase point operators which are used to describe the discrete Wigner function \cite{PhasePointOps}.

\section{Variants of the problem}

There are many different ways that one could define the `size' of the sets $\mathcal V_A,\mathcal V_B$. We will not consider specific measurements or specific types of generalised positivity in this work, so initially we will adopt an approach where we use various norms to define the size of an operator set. Our aim is to identify state spaces that could be useful in a broad range of situations, even though they may not be best choice for specific cases.

Ideally, the method we use to quantify the size of the sets $\mathcal V_A,\mathcal V_B$ should reflect how far from matrix-positive the operators within $\mathcal V_A,\mathcal V_B$ are. This is because if $\mathcal V_A,\mathcal V_B$ consist only of matrix-positive operators (which is of course not possible for an entangled state such as $\ket{\phi_d}$), then they will be in the dual of all quantum measurements and hence have a LHV model for all quantum measurements. If we restrict ourselves to sets $\mathcal V_A,\mathcal V_B$ of Hermitian operators, then the trace norm $ \|\cdot\|_1 $ captures this distance from matrix-positivity in a satisfying way. Indeed, for Hermitian operators $X$ the quantity $(\|X\|_{1} - \tr(X))/2 \geq 0$ is equal to the sum of the negative eigenvalues. If we further restrict our attention to local state spaces of only unit trace operators, then because tensor products of these operators will also be unit trace, and because the trace norm is multiplicative for tensor products of operators (i.e. $\|A\otimes B\|_1=\|A\|_1\|B\|_1$), the trace norm quantifies the non-positivity also for composite systems. This suggests that if we restrict our attention to sets $\mathcal V_A,\mathcal V_B$ of unit trace operators, then we could define the size of a single set $\mathcal S$ by
\begin{eqnarray}
\|\mathcal S\|_1 \defeq \sup \{ \|X\|_1 \mid X \in \mathcal S\},
\end{eqnarray}
and then define the size of both sets $\mathcal V_A,\mathcal V_B$ \textit{together} by the product of their individual sizes $\|\mathcal V_A\|_1 \|\mathcal V_B\|_1$.

However, finding smallest sets using the trace norm while incorporating the requirement of Hermiticity and unit trace appears to be difficult. So initially we will begin by analyzing a different problem where we abandon the condition of unit trace, and allow ourselves to consider more general norms $\|\cdot\|$ in place of the trace norm (so that now we define the size of a single set by $\|\mathcal S\|\defeq \sup \{ \|X\|\mid X \in \mathcal S\}$), although we consider the problem with or without the restriction of Hermiticity.

\bigskip

{\bf Problem 1:} For a quantum state $\Psi$ and for a suitable norm $\|\cdot\|$ what is the infimum {\it product size} $\|\mathcal V_A\| \|\mathcal V_B\|$ of all pairs of convex sets $(\mathcal V_A,\mathcal V_B)$ (with or without the Hermiticity constraint) such that $\Psi$ is $\mathcal V$-separable?

\bigskip

In section \ref{connections} we will find that (at least for multiplicative norms that satisfy the \emph{cross property} \cite{foot99}) this problem is equivalent to computing the so-called {\it projective tensor norm} (also known as {\it greatest cross norm}) that has been used already in the study of entanglement \cite{OR1,OR2}. This will enable us to draw on explicit formulas that have already been derived in previous works, as well as make strong connections to entanglement measures.

A drawback of Problem 1, whichever norm we use, is that there could be two choices $\mathcal V_A,\mathcal V_B$ and $\mathcal V'_A,\mathcal V'_B$ with the same size as measured by a norm, while still having (say) $\mathcal V_A \subset \mathcal V'_A$, and/or $\mathcal V_B \subset \mathcal V'_B$. In such cases, the sets $\mathcal V_A,\mathcal V_B$ will be the preferential choice, as under most notions of positivity if a set of operators is positive, then so must be any subset. So ideally we would like to know if there are {\it smallest} choices for $(\mathcal V_A,\mathcal V_B)$, in the following sense:

\bigskip

{\bf Problem 2:} Can we identify convex sets $(\mathcal V_A,\mathcal V_B)$ such that $\Psi$ is $(\mathcal V_A,\mathcal V_B)$-separable, and there exist no smaller sets $(\mathcal R_A,\mathcal R_B) \neq (\mathcal V_A,\mathcal V_B)$ with $\mathcal R_A \subseteq \mathcal V_A$ and $\mathcal R_B \subseteq \mathcal V_B$ such that $\Psi$ is $(\mathcal R_A,\mathcal R_B)$-separable?

\bigskip

In section \ref{smallest} will see that there are many such `smallest' solutions in the case of the maximally entangled state $\ket{\phi_d}\bra{\phi_d}$, and we will present methods of constructing a number of them. We will do this by initially tackling Problem 1 with the norm chosen to be the $ 2 $-norm, and then showing how the solutions to Problem 1 contain also solutions to Problem 2. In fact, in section \ref{incorporating} we will find that for the $ 2 $-norm and the maximally entangled state we are also able to incorporate additional requirements of strictly positive or unit trace quite straightforwardly---these conditions are useful both for meeting positivity requirements, and for technical reasons. In section \ref{prob3} we will used them for constructing solutions to another variant of the problem that we describe below. In Appendix~A we also present a solution of Problem 2 in the case of non-maximally entangled bipartite pure states, however, it does not incorporate the requirement of Hermiticity, and the product operators appearing in the separable decomposition are only in the dual of commuting local measurements.

In the context of constructing local hidden variable models for quantum states, there is yet another more natural variant of the problem involving convex cone state spaces \cite{Barnum}. The {\it conic hull} $\mathrm{conic}({\mathcal Y})$ generated by a set of operators $\mathcal Y$ is defined to be the smallest convex cone containing ${\mathcal Y}$, and can be generated by taking linear combinations of elements of ${\mathcal Y}$ with non-negative coefficients \cite{Boyd}. It is not difficult to see from the definition of the dual that if local state spaces $\mathcal V_A$ and $\mathcal V_B$ are contained in the dual of a collection of measurements $\mathcal F$, then so are the sets $\mathrm{conic}(\mathcal V_A)$ and $\mathrm{conic}(\mathcal V_B)$. Moreover, as $\mathcal V_A \subset \mathrm{conic}(\mathcal V_A)$ and $\mathcal V_B \subset \mathrm{conic}(\mathcal V_B)$, separability with respect to sets $\mathcal V_A$ and $\mathcal V_B$ implies separability with respect to the cones $\mathrm{conic}({\mathcal V_A})$ and $\mathrm{conic}({\mathcal V_B})$.
This means that if we are interested in using generalised separable decompositions to construct a local hidden variable models for as many quantum measurements as possible, then considering conic hulls can only increase the set of states that are separable, while not reducing the set of measurements for which separability implies the existence of a local hidden variable model. So our goal should be to look not for the smallest convex state spaces for which the quantum state is separable, but the smallest {\it convex cone} state spaces for which the quantum state is separable.

In this context it is important to note another issue that can arise. Consider one of the two subsystems, say $A$. Two convex cones $\mathcal T, \mathcal U$ of operators on $A$ satisfying $\mathcal T \subset \mathcal U$ can in principle be positive for the same set of quantum measurements. In such cases, separable decompositions involving these state spaces would give LHV models for the same situations, and there is no advantage in demonstrating separability using $ \mathcal T $ rather than $ \mathcal U $, even though $ \mathcal T $ is smaller. One way of dealing with this problem is to note that if $\mathcal Q_A$ is the set of local quantum states, then an operator $N$ gives $\tr(N \mathcal{Q_A})\geq 0$ iff $N$ is positive. This means that the set of quantum measurements for which a given conic hull $\mathrm{conic}(\mathcal V_A)$ is positive is fixed by the dual of $\mathrm{conic}({\mathcal V_A}\cup \mathcal Q_A)$. Hence to maximise the set of measurements for which generalised separable decompositions can supply a local hidden variable model, we should restrict our attention to local convex cone state spaces that contain the quantum states. This leads to the final variant of the problem that we consider:

\bigskip

{\bf Problem 3:} Can we identify local cones $(\mathcal V_A,\mathcal V_B)$ of operators that contain the local operators $\mathcal Q$, such that $\Psi$ is $(\mathcal V_A,\mathcal V_B)$-separable, and there exist no smaller cones $(\mathcal R_A,\mathcal R_B) \neq (\mathcal V_A,\mathcal V_B)$ with $\mathcal R_A \subseteq \mathcal V_A$ and $\mathcal R_B \subseteq \mathcal V_B$ with these properties?

\bigskip

Any two different solutions to Problem 3 will be the duals of distinct sets of quantum measurements, and hence provide local hidden variable models for different scenarios.
In section \ref{prob3} we will see that some of our solutions to Problems 1 and 2 for the maximally entangled state also enable us to construct solutions to Problem 3.

\section{Summary of results and method} \label{summ}

The logical structure of the arguments that we use to solve these problems is as follows:
\begin{itemize}
\item We will first show that the solution to problem 1 is equivalent to computing a particular kind of norm (known as a {\it cross norm}) for the quantum state under consideration.
\item We then show that for sets of local operators of exactly this size, a separable decomposition is only possible if the sets contain at a minimum number of operators of a certain minimal size (as measured by the 2-norm).
\item In the case of maximally entangled states we present examples of sets of operators that are the convex hull of such minimal sets, and argue that no strict subset of them can be chosen while keeping the state generalised separable, because any smaller subset contains too few operators of the minimal required size. These sets are hence solutions to Problem 2. The argument is generalised to arbitrary bipartite pure states in Appendix ~A.
\item In the case of the maximally entangled state we show that such sets can be chosen to have unit trace (including subsets of phase point operators defining discrete Wigner functions as a particular case).
\item We then show that the conic hulls generated by these unit trace convex sets provide solutions to Problem 3.
\end{itemize}

\section{Connections to cross norms}\label{connections}

In this section we show that Problem 1 is closely related to the notion of cross norms. In particular we show that the cross norm $\|\Psi\|_{\gamma}$ (defined below) is exactly the minimum possible value of $\|\mathcal V_A\|\|\mathcal V_B\|$ in Problem 1.

Consider any norm $\|\cdot\|$ such that on two state spaces it satisfies the cross property \cite{foot99}.
For instance, the well-studied family of Schatten $p$-norms $\|X\|_p = \tr(|X|^p)^{1/p}$, where $1 \leq p < \infty$, obey the cross property. For any fixed norm $\| \cdot \|$ on a state space we consider the projective tensor norm, denoted by $\|\cdot\|_\gamma$, on the tensor product of two spaces:
\begin{eqnarray}
\|X\|_\gamma \defeq \inf\left\{\sum\nolimits_{i}\|A_i\|\|B_i\| \mid X = \sum\nolimits_{i} A_i \otimes B_i \right\}, \label{gcn}
\end{eqnarray}
where the infimum is over finite sums of arbitrary operators $A_i, B_i$ (not necessarily in $\mathcal V_A$ or $\mathcal V_B$ respectively). If the norm $\| \cdot \|$ inside the sum is equal to the $p$-norm $\| \cdot \|_p$, then we denote the corresponding projective norm by  $\| \cdot \|_{\gamma, p}$. It has been shown by one of us \cite{OR1} that a bipartite quantum state $\rho$ is quantum separable if and only if $\|\rho\|_{\gamma, 1} = 1$. Moreover, the cases of $\|\cdot\|_{\gamma, 1}$ and $\|\cdot\|_{\gamma, 2}$ have been used in the definition of entanglement measures \cite{OR1}, while $\|\cdot\|_{\gamma, 2}$ was also used in \cite{OR2} to formulate a computable separability criterion.

Now let $\Psi$ be a bipartite quantum state (in this section we do not need to assume that it is $\ket{\phi_d}\bra{\phi_d}$) and $\| \cdot \|$ be a (fixed) norm on the local state spaces. If $\Psi$ is $\mathcal V$-separable, then
\begin{eqnarray}
\Psi =\sum_{k} p_k A_k \otimes B_k, \quad A_k \in \mathcal V_A, \,\, B_k \in \mathcal V_B,
\end{eqnarray}
for some choices of operators $A_k,B_k$. From the definition of the cross norm in Eq. (\ref{gcn}),  it follows that
\begin{eqnarray}
\|\Psi\|_{\gamma} \leq \sum_{k} p_k \|A_k\|\|B_k\| \leq \max_k \|A_k\|\|B_k\|\leq \|\mathcal V_A\|\|\mathcal V_B\|.\label{oneway}
\end{eqnarray}
Hence, we see that the infimum product size of the local sets is lower bounded by the corresponding cross norm. We now show that the opposite inequality holds. For simplicity we assume that infimum of Eq. (\ref{gcn}) is an achievable minimum (the argument can be modified to hold even when the infimum is not achievable). Under this assumption let
\begin{eqnarray}
\Psi = \sum_{i} A_i \otimes B_i
\end{eqnarray}
be the finite decomposition of $\Psi$ that achieves $\|\Psi\|_{\gamma}$ in Eq. (\ref{gcn}). We may define a probability distribution by $p_i \defeq  \|A_i\| \|B_i\| /\|\Psi \|_{\gamma}$. Then we have that:
\begin{eqnarray}
\Psi = \sum_{i} p_i \frac{\sqrt{\|\Psi\|_{\gamma}}}{\|A_i\|} A_i \otimes
\frac{\sqrt{\|\Psi\|_{\gamma}}}{\|B_i\|}B_i. \label{sepdeco}
\end{eqnarray}
Now let $\mathcal V_A'$ be the convex hull of the set of all operators $\sqrt{\|\Psi\|_{\gamma}} A_i / \|A_i\|$ and similarly let $\mathcal V_B'$ be the convex hull of the set of all operators $ \sqrt{\|\Psi\|_{\gamma}}B_i/\|B_i\|$. In the separable decomposition (\ref{sepdeco}), all the local operators in the product terms from the sets $(\mathcal V_A', \mathcal V_B')$ have the same norm, equal to $\sqrt{\|\Psi\|_{\gamma}}$. The triangle inequality hence implies that $\|\mathcal V_A'\| = \|\mathcal V_B'\| = \sqrt{\|\Psi\|_{\gamma}}$, and hence that $\|\mathcal V_A'\| \|\mathcal V_B'\| = \|\Psi\|_{\gamma}$. Hence, there are choices of state spaces $\mathcal V_A',V_B'$ such $\|\mathcal V_A'\| \|\mathcal V_B'\| = \|\Psi\|_{\gamma}$, and the inequality (\ref{oneway}) is tight.

This connection allows us to use existing results on the calculation of the cross norm. In particular, the value of $\|\cdot\|_{\gamma, 1}$ and $\|\cdot\|_{\gamma, 2}$ has been calculated for (among others) all pure bipartite states \cite{OR1}, with or without the requirement of Hermiticity.

We will now build upon these results to provide solutions to Problem 2. In the next section we will begin this analysis by rederiving some of the results of \cite{OR1} for {$\|\cdot\|_{\gamma, 2}$} in the case of interest to us (the maximally entangled state). We will use these observations to provide a variety of optimal solutions $\mathcal V_A,\mathcal V_B$, and then also provide solutions to Problem 2.

\section{Solutions to Problem 2 for maximally entangled states}\label{smallest}

We begin by expanding a $d\times d$ matrix of a single system operator $X \in \mathcal S$ in an orthogonal basis of $d^2$ Hermitian operators $C_i$:
\begin{eqnarray}
X = \sum_{i=1}^{d^2} x_i C_i,
\end{eqnarray}
where the expansion coefficients $x_i\in\mathbb{C}$ form a $ d^2 $-dimensional vector $\boldsymbol{x}\defeq(x_1,x_2,\dots,x_{d^2})$, and the operator basis is chosen to satisfy the condition $\mathrm{tr}(C_iC_j)=d~\delta_{ij}$. An example of such a basis for qubit systems is the set of Pauli operators with the identity. In such an expansion, the square of the 2-norm of the operator $X$ is given by $\tr(X X^{\dag})=\sum_{i,j}x_i x^*_j\tr(C_i C_j)=d\sum_{i,j}x_i x^*_j\delta_{i,j}=d~\|\boldsymbol{x}\|^2$, hence {
\begin{eqnarray}
\label{2norm_S}
\|X\|_2 = \sqrt{d}\|\boldsymbol{x}\|,
\end{eqnarray}
where $\|\boldsymbol{x}\|$ is the standard Euclidean norm of the vector $\boldsymbol{x}$.

{\bf Theorem 1} Consider any convex sets of operators $\mathcal V_A, \mathcal V_B$ for which $\ket{\phi_d}$ is $(\mathcal V_A,\mathcal V_B)$-separable. Then the following must hold: (a) $\mathcal V_A,\mathcal V_B$ must satisfy $\|\mathcal V_A\|_2\|\mathcal V_B\|_2 \geq d$; if $\|\mathcal V_A\|_2\|\mathcal V_B\|_2 = d$ then (b) the separable decomposition must involve only product terms $A_k \otimes B_k$ with $\|A_k\|_2\|B_k\|_2 = d$, and (c) a separable decomposition must contain at least $d^2$ operators from $\mathcal V_A$ and $\mathcal V_B$ each. Finally, let $\{C_i\}_{i=1}^{d^2}$ be any orthogonal basis of Hermitian operators for $d\times d$ matrices such that each $C_i$ has 2-norm $ \|\mathcal C_i\|_2=\sqrt{d} $, if $\mathcal V_A$ is chosen to be the convex hull of the $C_i$ and $\mathcal V_B$ is chosen to be the convex hull of their transpositions $C_i^T$, then (d) $\ket{\phi_d}$ is $(\mathcal V_A,\mathcal V_B)$-separable, and (e) $\ket{\phi_d}$ is inseparable with respect to to any strict subsets of $\mathcal V_A,\mathcal V_B$.

{\bf Proof:} Parts (a) and (c) are existing results, either following from the calculation of $\|\cdot\|_{\gamma, 2}$ in \cite{OR1}, or from the Schmidt decomposition applied to the operator space. However, we will also rederive them as it will help us to prove the remaining parts. First, note that if there is a convex operator decomposition such as:
\begin{eqnarray}
\ket{\phi_d}\bra{\phi_d} = \sum_k p_k A_k \otimes B_k,
\end{eqnarray}
where $p_k$ is a probability distribution and $A_k \in \mathcal V_A$ and $B_k \in \mathcal V_B$, then the Hilbert-Schmidt inner product of both sides with the basis of Hermitian operators $C_i \otimes C^T_j$ must match. Expressing $A_k = \sum {\alpha^k_i C_i}$ and $B_k = \sum {(\beta^k_j)^* C^T_j}$, where ${\alpha^k}$ and $({\beta^k})^*$ (the conjugate is incorporated into the definition for later convenience) are complex expansion vectors representing $A_k$ and $B_k$, respectively, and using $\tr(C_iC_j) =\tr(C^T_iC^T_j) =d~\delta_{ij}$ and the identity $\bra{\phi_d}X \otimes Y\ket{\phi_d}=\tr(XY^T)/d$, we must have:
\begin{eqnarray}
\delta_{ij} = d^2 \sum_k p_k {\alpha^k_i} ({\beta^k_j})^*. \label{match}
\end{eqnarray}
All the statements of the theorem are short consequences of the above identity. In particular, summing over the $d^2$ terms involving $i=j$ gives:
\begin{eqnarray}
d^2 &=& d^2 \sum_k p_k \sum_i \alpha^k_i (\beta^k_i)^*, \nonumber \\
\Rightarrow 1 &=& \sum_k p_k \langle \boldsymbol{\beta}^k,\boldsymbol{\alpha}^k\rangle,
\label{vectorsep}
\end{eqnarray}
where $\langle \boldsymbol{\beta}^k,\boldsymbol{\alpha}^k\rangle$ represents the inner product. This means that the average of the inner products of $\boldsymbol{\alpha}^k$ and $\boldsymbol\beta^k$ is equal to 1. Hence, by convexity and the Cauchy-Schwarz inequality, it must be the case that $\max_k\|\boldsymbol\alpha^k\|\|\boldsymbol\beta^k\| \geq 1$, and hence, using Eq. \eqref{2norm_S} and the fact that $\|\mathcal{S}\|$ is no less than $\|\cdot\|$ for one of its elements, gives $\|\mathcal V_A\|_2\|\mathcal V_B\|_2 \geq d$, proving (a).
If we now restrict our attention to only sets satisfying $\|\mathcal V_A\|_2\|\mathcal V_B\|_2 = d$, this means that we must have $\|\boldsymbol\alpha^k\|\|\boldsymbol\beta^k\|\leq 1$. Hence, by convexity and the Cauchy-Schwarz inequality, the only way that Eq. (\ref{vectorsep}) can be true is if $\|\boldsymbol\beta^k\|^2 \boldsymbol\alpha^k = \boldsymbol\beta^k$, and hence $\langle \boldsymbol\beta^k,\boldsymbol\alpha^k\rangle=1$ for all $k$. This implies (b), and shows that there is a trade-off---the smaller the $ 2 $-norm of one state space, the larger must be the $ 2 $-norm of the other.

To see that at least $d^2$ operators are required, let us put the fact that $\|\boldsymbol\beta^k\|^2 \boldsymbol\alpha^k = \boldsymbol\beta^k$ back into Eq. (\ref{match}) to get
\begin{eqnarray}
\delta_{i,j} &=& d^2 \sum_k p_k \|\boldsymbol\beta^k\|^2 \alpha^k_i (\alpha^k_j)^*, \nonumber \\
\Rightarrow {\delta_{i,j} \over d^2} &=& \sum_k \tilde{\alpha}^k_i (\tilde{\alpha}^k_j)^*, \label{match2}
\end{eqnarray}
where we have defined new unnormalized vectors $\tilde{\boldsymbol\alpha}^k \defeq \sqrt{p_k}\|\boldsymbol\beta^k\|\boldsymbol\alpha^k$. We may now reinterpret this equation in the following way. For fixed $i$ we consider the coefficients $\tilde{\alpha}^k_i$ for varying $k$ to be coefficients of a vector of length $N$, where $N$ (which in principle could be very large) is the number of different values of $k$ in the sum (\ref{match2}). Then Eq. (\ref{match2}) tells us that we have $d^2$ such vectors of norm $1/d$, and they form an orthogonal set (in the dimension $N$ vector space). For it to be possible to pick $d^2$ orthogonal vectors, $k$ must range over at least $d^2$ values, hence proving (c).

To show (d) note that setting $p_k=1/d^2$, ${\beta}^k_j \defeq \delta_{jk}$, and ${\alpha}^k_j \defeq \delta_{jk}$,
trivially satisfies equation (\ref{match}), showing that (the convex hull of) any orthogonal basis $\{ C_i \}$ of operators satisfying $\tr(C_i C_j)=d~\delta_{ij}$ provides suitable choices for $\mathcal V_A$ and $\mathcal V_B$ (by setting $\mathcal V_B=\mathcal V^T_A$).

Finally to show (e) note that for $\lambda\in(0,1)$ the {\it strict} inequality $\|\lambda X + (1-\lambda)Y\|_2 < \|\lambda X\|_2+\|(1-\lambda)Y\|_2 \leq \max\{\|X\|_2,\|Y\|_2\}$ holds if $X$,$Y$ are not proportional to each other (this follows from Cauchy-Schwarz, and it does not hold for trace norm). Hence, no other operators within the convex hull of the $d^2$ operators $C_i$ can attain a 2-norm of $\sqrt{d}$, and hence a strict subset cannot satisfy the necessary condition (b).
$\blacksquare$

These observations provide us with a method of constructing solutions to Problems 1 and 2. However, in some contexts it is useful to make further restrictions, e.g. that the operators in $\mathcal V_A,\mathcal V_B$ are of strictly positive trace or of unit trace. It is straightforward (we describe this in the next section) to include these requirements in the context of the 2-norm. We will use the resulting solutions to construct solutions to Problem 3.

In Appendix ~A, we prove that parts of Theorem 1 also generalize to arbitrary bipartite pure states, providing one solution of Problem 2 for all bipartite pure states.

\section{Incorporating positive or unit trace}\label{incorporating}

\begin{figure}
\includegraphics[width=.5\textwidth]{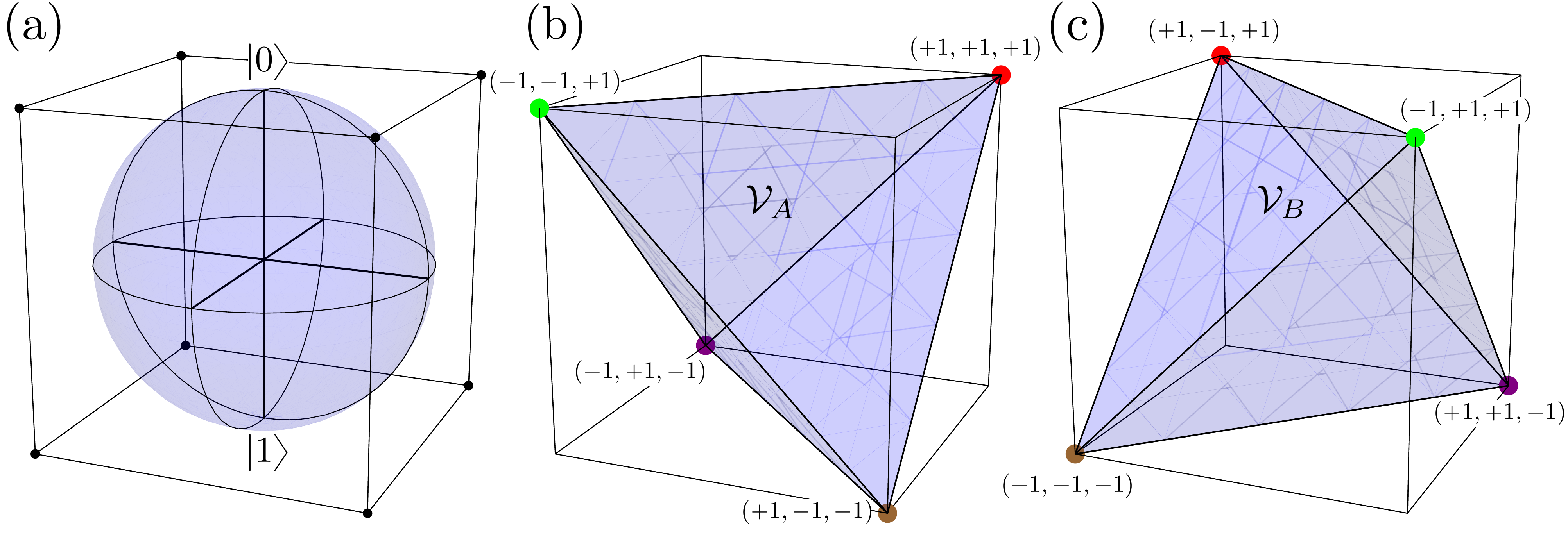} \centering
\caption{\label{fig:lattice} (a) The Bloch ball representation of a qubit state $ \rho(\boldsymbol{x})=\frac{1}{2}(\mathbb{1}+\boldsymbol x.\boldsymbol{\sigma}) $, where $ \boldsymbol{x} $ is a three dimensional real vector, and $ \boldsymbol{\sigma} $ is a vector of the three qubit Pauli operators. The eight vertices of the cube correspond to non-quantum operators with $ \boldsymbol{x}=(\pm 1, \pm 1, \pm 1) $.  (b) and (c) are the two sets $ \mathcal{V}_{A} $ and $ \mathcal{V}_{B} $ of local operators on systems A and B, respectively, with each set containing four vertices of the cube satisfying the condition $ \tr(C_iC_j)=2\delta_{ij} $. Specifically, the qubit maximally entangled state is separable with respects to these two sets, such that $ \ket{\phi_2}\bra{\phi_2}=\frac{1}{4}(\rho(1,1,1)\otimes\rho(1,-1,1)+\rho(-1,-1,1)\otimes\rho(-1,1,1)+\rho(1,-1,-1)\otimes\rho(1,1,-1)+\rho(-1,1,-1)\otimes\rho(-1,-1,-1))$.
\label{fig:tet}
}
\end{figure}

The previous sections show that any orthogonal basis of $d^2$ Hermitian operators normalized to $\tr(C_iC_j)=d\ \delta_{ij}$ provides a solution to both Problem 1 (in the case of the $ 2 $-norm) and Problem 2. However, these solutions can in principle contain operators that aren't positive for some important forms of generalised positivity. For instance, if an operator has negative or complex trace, then it cannot be in the dual of any POVM. So we would like to consider adding a constraint that the trace is positive. In this section we show how this can be done straightforwardly for maximally entangled states.

To obtain such solutions we simply need to find bases of Hermitian operators with positive trace. This can be done using the Gram-Schmidt process. If we start from any Hermitian basis $C_i$ for which the first element is the identity $C_1 = \mathbb{1}$, and the remaining $C_i$ are traceless, then imposing the requirement that $\mathcal V_A,\mathcal V_B$ contain Hermitian operators with positive trace amounts to demanding that the expansion vectors $\boldsymbol\alpha^k,\boldsymbol\beta^k$ are real, and that their first components are positive numbers. If we consider only solutions that are constructed as the convex hull of $d^2$ orthogonal operators of $ 2 $-norm $\sqrt{d}$, then finding the appropriate vectors $\boldsymbol\alpha^k,\boldsymbol\beta^k$ is equivalent to picking a $d^2\times d^2$ real orthogonal matrix such that the top row consists of real positive coefficients, and can be solved using the Gram-Schmidt procedure.

In the next section we will need to use solutions that are not only of positive trace, but of unit trace. We can obtain such solutions in the same way: if we consider only solutions that are constructed as the convex hull of $d^2$ orthogonal operators of $ 2 $-norm $\sqrt{d}$, then finding the appropriate vectors $\boldsymbol\alpha^k,\boldsymbol\beta^k$ is equivalent to picking a $d^2\times d^2$ real orthogonal matrix such that the top row consists of $(1/d,1/d,1/d,..)$, and this can also be solved using the Gram-Schmidt procedure.

Amongst these unit trace solutions there exists one type that is already widely used in the construction of classical models: the $d^2$ subsets of the {\it phase point operators} \cite{PhasePointOps} that describe discrete Wigner functions. Each such subset provides a Hermitian unit trace orthogonal basis of the correct norm. In the case of $d=2$ it can be shown that the only unit trace Hermitian operator basis satisfying $\tr(C_iC_j)=d~\delta_{ij}$ are tetrahedra that unitary rotations or transpositions of the example presented in Fig. \ref{fig:tet}. However, for higher dimensions $d>2$, there are inequivalent solutions that do not share the same spectrum and hence are not unitarily equivalent to subsets of phase point operators. Our analysis shows that any quantum measurements in the dual of such sets will have local hidden variable models for the maximally entangled state, going beyond constructions available via a discrete Wigner function approach.

\section{Solutions to Problem 3 for the maximally entangled state}\label{prob3}

In this section we show that conic hulls generated from the unit trace operator bases in the previous section can enable us to provide solutions to Problem 3. The strategy of our solutions to Problems 1 and 2 was to show that if the maximally entangled state is separable with respect to given state spaces, then there must be operators in those state spaces of a big enough norm. As convex cones contain operators of arbitrary norm, we cannot apply this strategy to Problem 3 without modification. We will get around this problem by restricting our attention to cones that can be generated as the conic hulls of convex sets of operators with unit trace. We will argue that these convex sets cannot be made smaller while preserving separability, and thereby also argue that the convex cones cannot be made smaller while preserving separability.

Let ${\mathcal W_A},{\mathcal W_B}$ denote any set of unit trace orthogonal basis operators constructed in the previous section, and let ${\mathcal Q_A},{\mathcal Q_B}$ denote the local quantum states on systems $A,B$ respectively. Consider $\mathrm{conic}({\mathcal W_A}\cup \mathcal Q_A)$ and $\mathrm{conic}({\mathcal W_B}\cup \mathcal Q_B)$. As the generators of these conic hulls all have strictly positive trace, all operators in the resulting cones will have strictly positive trace except for the $\mathbf{0}$ operator. This means that we can assume that any cone-separable decomposition of the maximally entangled state:
\begin{eqnarray}
\ket{\phi_d}\bra{\phi_d} =\sum_{k} p_k A_k \otimes B_k, \quad A_k \in \mathrm{conic}({\mathcal W_A}\cup \mathcal Q_A), \,\, B_k \in \mathrm{conic}({\mathcal W_B}\cup \mathcal Q_B) \nonumber
\end{eqnarray}
only contains local operators on the right hand side that have strictly positive trace (any contribution from the trivial $\mathbf{0}$ operator can be discarded). Hence by dividing the operators on the right hand side by their trace, we may recover a separable decomposition in terms of only {\it unit trace} operators from the cones:
\begin{eqnarray}
\ket{\phi_d}\bra{\phi_d} =\sum_{k} p_k \tr (A_k) \tr (B_k) {A_k \over \tr (A_k)} \otimes {B_k \over \tr (B_k)}. \label{unittrace}
\end{eqnarray}
This means that if a state is separable with respect to the conic hulls of operators with strictly positive trace, then the state is also separable with respect to the convex subsets of the cones consisting of only unit trace operators.

To these convex subsets we may now apply Theorem 1. In order for such a separable decomposition (\ref{unittrace}) to exist for the maximally entangled states, then we know that the unit trace operators on the r.h.s. must have a minimum 2-norm of $\sqrt{d}$.
Note that any unit trace operator in the conic hulls must be a convex combination of the unit trace generators. Using the fact that for the 2-norm the {\it strict} inequality $\|\lambda X + (1-\lambda)Y\|_2 < \|\lambda X\|_2+\|(1-\lambda)Y\|_2 \leq \max\{\|X\|_2,\|Y\|_2\}$ holds if $X$,$Y$ are not proportional to each other, we hence see that the operators appearing in (\ref{unittrace}) must be precisely the operators from ${\mathcal W_A},{\mathcal W_B}$, because the quantum states have a 2-norm that is too small (their 2-norm $\leq 1$). The only unit trace member of the conic hulls with a high enough  2-norm of $\sqrt{d}$ are hence the ${\mathcal W_A},{\mathcal W_B}$---all the extremal points of the original sets $ \mathcal{W} $ are needed for the separable decomposition because all other unit trace operators have a 2-norm that is strictly less than $\sqrt{d}$, thereby violating the requirement of Theorem 1 part (b). Hence the conic hull state spaces cannot be made smaller, as they must contain the ${\mathcal W_A},{\mathcal W_B}$  as well as the local quantum states. Hence we have the following:

{\bf Theorem 2} Consider $\mathrm{conic}({\mathcal W_A}\cup \mathcal Q_A)$ and $\mathrm{conic}({\mathcal W_B}\cup \mathcal Q_B)$ where ${\mathcal W_A},{\mathcal W_B}$
give a unit trace solution to Problem 2 for the maximally entangled state. Then $\mathrm{conic}({\mathcal W_A}\cup \mathcal Q_A)$ and $\mathrm{conic}({\mathcal W_B}\cup \mathcal Q_B)$ is solution to Problem 3 for the maximally entangled state.

\begin{figure}
\includegraphics[width=.5\textwidth]{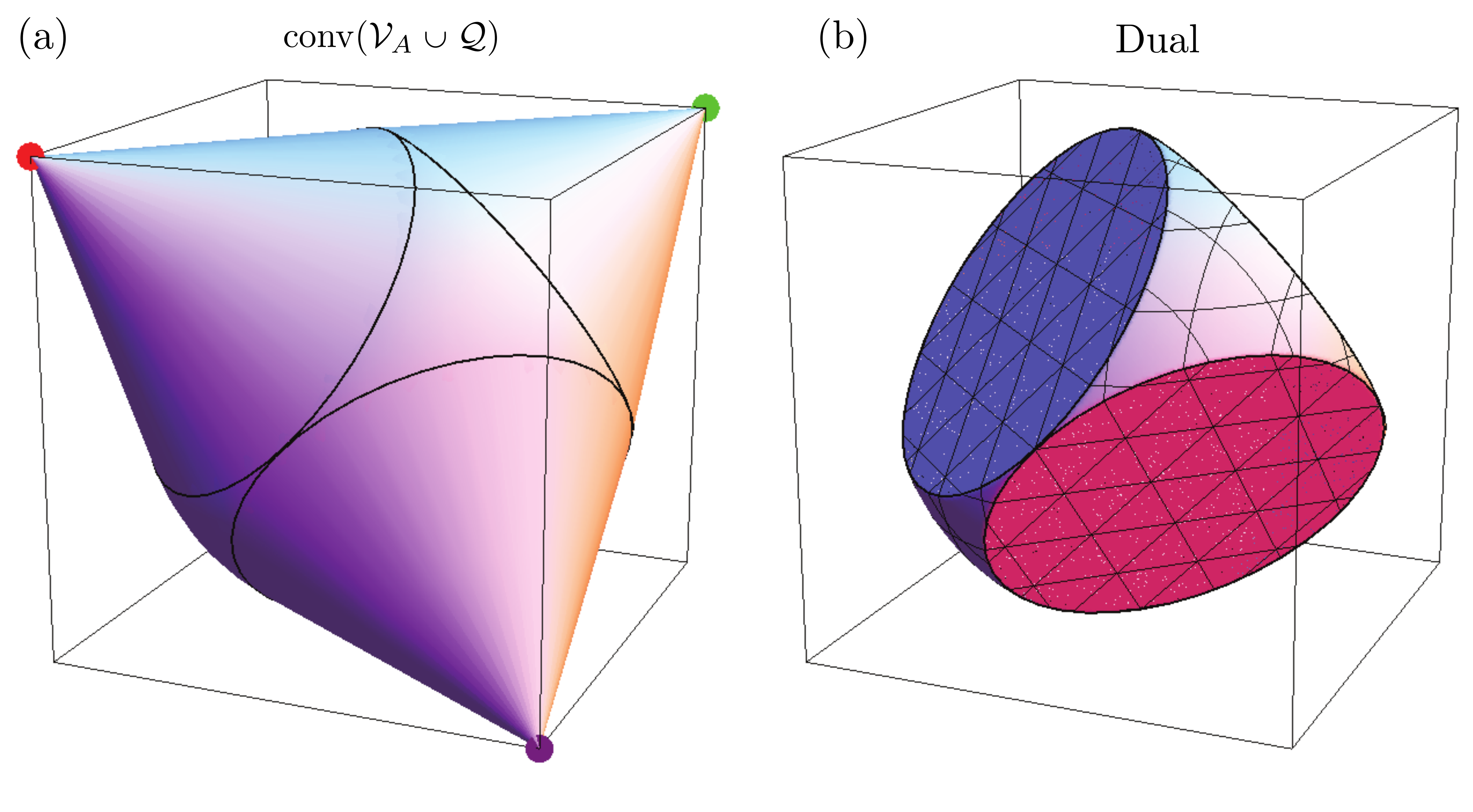} \centering
\caption{\label{fig:lattice} Generators of conic hulls for Problem 3 when $d=2$. (a) The convex hull the quantum states (Bloch sphere) with the set $ \mathcal{V}_A $ shown in Fig. 1(b). The convex cone generated by this set features in a solution to Problem 3 (b) The dual of the convex hull, representing the quantum measurement directions for which $ \mathcal{V}_A $ is in the dual.
\label{fig:conv}
}
\end{figure}
This implies that if we consider any two unitarily inequivalent $ {\mathcal W}= ({\mathcal W_A,W_B}) $ and $ {\mathcal W'}= ({\mathcal W'_A,W'_B})  $ that have been constructed for Problem 2, as is possible for $d>2$, then the separable decomposition resulting from the conic hulls of these states with the quantum states will supply LHV models for distinct and unitarily inequivalent sets of measurements. In this sense our constructions generalise the local hidden variable models that one can construct from discrete Wigner functions.

Fig. \ref{fig:conv} depicts (a) the convex hull of ${\mathcal W_A}$ and ${\mathcal Q_A}$, and (b) its dual, when $d=2$. Operationally this means that the qubit Bell state $\ket{\phi^+}$ has a local hidden variable model for all POVMs $M_A$ on A and $M_B = M_A^T$ on B, where the elements of $M_A$ are proportional to $\mathbb{1} + \boldsymbol r\cdot \boldsymbol\sigma$ and $\boldsymbol r$ is a vector from the convex set in Fig. \ref{fig:conv}(b).

\section{Conclusions}\label{conclusions}

We have determined local state spaces that admit a separable decomposition of an entangled pure state $\ket{\psi}$ and cannot be made strictly smaller while maintaining separability. In the context of maximally entangled states, in particular where the local state spaces can be chosen to have unit trace, this has applications in constructing local hidden variable models.

Our measure of `smallest' state space is given by the operator 2-norm not only because it renders the optimization of Problem 1 analytically tractable, but also because it enables solutions of Problem 2 and Problem 3. We do note, however, that using the trace norm would be more natural when searching for states spaces of operators that are not very negative; further work is required to explore this option.

We have made a connection between cross norms and generalized separability, and it is likely that these connections can be generalized when considering other notions of positivity for the local state spaces.

It will also be interesting to know whether it is possible to extend the method from the bipartite to the multipartite case, where very little is known about classical models for quantum states.

\ack
The authors would like to thank David Gross, Marco Piani, and Terry Rudolph for fruitful discussions. HA, SJ and SV are supported by EPSRC grant EP/K022512/1.

\appendix

\section{A solution to Problem 2 for bipartite pure states}

We may obtain to a solution to Problem 2 for general bipartite pure states by applying similar considerations to in the maximally entangled case. Consider a bipartite pure state written in Schmidt form:
\begin{eqnarray*}
\ket{\psi} = \sum_i \lambda_i \ket{ii}.
\end{eqnarray*}
We assume that the Schmidt rank is maximal, else we truncate the local quantum state spaces to dimension $d$, where $d$ is the Schmidt rank. Using the Schmidt basis $\ket{i}$, we may construct an orthogonal basis for the local operator space consisting of the $d^2$ operators:
\begin{eqnarray*}
C_{ij} = \sqrt{d} \ket{i}\bra{j},
\end{eqnarray*}
where $d$ is the Schmidt rank of $\ket{\psi}$. Note that $\tr(C_{ij}C^T_{kl})=d \ \delta_{ik}\delta_{jl}$.
Suppose that we have a separable decomposition of $\ket{\psi}$ as follows:
\begin{eqnarray*}
\ket{\psi}\bra{\psi} = \sum^N_{k=1} p_k A_k \otimes B_k.
\end{eqnarray*}
Let us decompose the local operators as $A_k=\sum_{ij} \alpha^k_{ij}C_{ij}$ and
$B_k=\sum_{ij} (\beta^k_{ij})^*C^T_{ij}$. Then we may compute:
\begin{eqnarray}
\bra{\psi} C_{gh} \otimes C_{ij} \ket{\psi} & = & d \ \lambda_g \lambda_h \delta_{gi}\delta_{hj},\notag\\
 &=& d^2 \sum^N_{k=1} p_k \alpha^k_{gh} (\beta^k_{ij})^*. \label{match3}
\end{eqnarray}
Summing the equation over $g,h,i,j$ gives:
\begin{eqnarray}
{(\sum_g \lambda_g)^2 \over d} = \sum_k p_k \langle \boldsymbol\beta^k, \boldsymbol\alpha^k\rangle, \nonumber
\end{eqnarray}
where $\boldsymbol\alpha^k$ and $\boldsymbol\beta^k$ are $d^2\times 1$ vectors of coefficients $\alpha^k_{ij}$ and $\beta^k_{ij}$ respectively, and $\langle \boldsymbol\beta^k, \boldsymbol\alpha^k\rangle$ symbolizes their inner product. If we make the restriction that the local operators in the separable decomposition are of 2-norm $\leq (\sum_g \lambda_g)^2$, then the only way that this equation can be true is if each $\boldsymbol\alpha^k$ and $\boldsymbol\beta^k$ are proportional, and have product 2-norm equal to $(\sum_g \lambda_g)^2/d$. We may place this finding back into equation (\ref{match3}) to get:
\begin{eqnarray}
\bra{\psi} C_{gh} \otimes C_{ij} \ket{\psi} &=& d \ \lambda_g \lambda_h \delta_{gi}\delta_{hj},\notag\\
&=& d^2 \sum^N_{k=1} p_k \alpha^k_{gh} (\alpha^k_{ij})^*. \nonumber
\end{eqnarray}
This equation may be reinterpreted as an orthogonality relation between $d^2$ vectors labeled by $i,j$, with $N$ components in each vector (the size of the range of $k$).
Hence if we need at least $N \geq d^2$ orthogonal vectors of dimension $d^2$. If we set $N=d^2$ then we may arrive at a solution by setting $p_k=1/d^2$, and then picking an orthogonal set of vectors normalized such that the vector labeled by $i,j$ has Euclidean norm $\sqrt{d\lambda_i\lambda_j}$. As with the maximally entangled case, any such solution is a smallest one, as no other set of $d^2$ operators in the convex hull can have a high enough 2-norm.

One disadvantage of this solution is that some of the operators in the local state spaces are only positive for measurements that are diagonal in the computational basis (as positive numbers summing to zero must themselves be zero, the fact that $C_{ij}$ are traceless for $i \neq j$ implies that $\bra{j}M\ket{i}=0$ for any measurement element $M$ for which $C_{ij}$ is positive). However, there are other notions of positivity (particularly in scenarios involving post-selection, see e.g. \cite{AJRV2}) that these local state spaces can nevertheless have.


\begin{thebibliography}{70}
\bibitem{foot98} For convenience we explain this argument here. Suppose that
\begin{equation}
\rho = \sum_i p_i A_i \otimes B_i \nonumber
\end{equation}
is a separable decomposition of normalised operator $\rho$, where $A_i$ and $B_i$ are $\mathcal{F}$-positive (for simplicity we assume that the same measurements are available to both parties). Then for local POVMS $\mathcal M$ and $\mathcal N$ the outcome probabilities are given by
\begin{eqnarray}
\tr (\rho M_k \otimes N_l) &=& \sum_i p_i \tr (A_i M_k) \tr (B_i N_l), \nonumber \\
&=& \sum_i p_i a(k|i) b(l|i), \label{lhv}
\end{eqnarray}
where $a(k|i):= \tr (A_i M_k)$ and $b(l|i):=\tr (B_i N_l)$. Now we must have that $\sum_k a(k|i) = \tr (A_i) $ and $\sum_l b(l|i) = \tr (B_i)$ from the completeness of the local POVMs, and $a(k|i),b(l|i)\geq 0$ from the fact that $A_i,B_i$ are $\mathcal{F}$-positive. If $A_i$ or $B_i$ are traceless, then
that means that the $a(k|i)$ or $b(l|i)$ will be all zero, and will not contribute to the sum in equation (\ref{lhv}), so we may rewrite equation (\ref{lhv}) as:
\begin{eqnarray}
\tr(\rho M_k \otimes N_l) &=& \sum_{\{i| \tr(A_i),\tr (B_i)>0\} } p_i a(k|i) b(l|i), \nonumber \\
&=& \sum_{\{i| \tr (A_i),\tr (B_i)>0\} } p_i \tr (A_i) \tr (B_i)  {a(k|i) \over \tr(A_i)  } {b(l|i) \over \tr(B_i) }. \label{lhv2}
\end{eqnarray}
Defining the normalised probability distribution $q_i$ and the normalised conditional probability distributions $r(k|i)$ and $s(l|i)$ by:
\begin{eqnarray}
q_i &&:= p_i \tr (A_i) \tr(B_i), \nonumber \\
r(k|i)&&:={a(k|i) \over \tr (A_i) },  \nonumber \\
s(l|i)&&:={b(l|i) \over \tr(B_i) }, \nonumber
\end{eqnarray}
(the normalisation of $q_i$ is enforced by the fact that $\rho$ has unit trace), we see that equation (\ref{lhv2}) supplies a local hidden variable model for the probability distribution.
\bibitem{foot99}A norm is said to satisfy the cross property if $\|A\otimes B\| =\|A\| \|B\|$.
\end{thebibliography}
\end{document}